\documentclass{pasj00}
\begin{document}
\SetRunningHead{S.\,G.\,Ryu et al.}{An X-ray Face-on View of the Sgr B Molecular Clouds}
\Received{2009/01/26}
\Accepted{2009/04/20}
\author{Syukyo G. \textsc{Ryu}, Katsuji \textsc{Koyama},\\ Masayoshi \textsc{Nobukawa}, Ryosuke \textsc{Fukuoka}, and Takeshi Go \textsc{Tsuru}}  
\affil{Department of Physics, Graduate School of Science, Kyoto University, Sakyo-ku, Kyoto 606-8502}
\email{ryu@cr.scphys.kyoto-u.ac.jp}
\title{An X-Ray Face-on View of the Sgr B Molecular Clouds \\ Observed with Suzaku}
\KeyWords{Galaxy: center ---  ISM: molecules, clouds, structure --- X-ray spectra} 
\maketitle

\begin{abstract}
We present a new methodology to derive the positions of the Sgr B molecular clouds (MCs) along the line of sight, as an application study of the Galactic center diffuse X-rays (GCDX). The GCDX is composed of hot plasma emission of about 7 keV and 1 keV temperatures, and non-thermal continuum emission including the 6.4 keV line from neutral irons. The former, the Galactic center plasma emission (GCPE), is uniformly distributed over 1 degree in longitude, while the latter is clumpy emission produced by Thomson scattering and fluorescence from MCs irradiated by external X-rays (the X-ray reflection nebula emission: XRNE). We examined the Suzaku X-ray spectra of the GCPE and XRNE near to the Sgr B MC complex, and found that the spectra suffer from two different absorptions of $N_{\rm H}$ (Abs1) $\geq 10^{23}$~H~cm$^{-2}$ and $N_{\rm H}$ (Abs2) $\simeq 6\times$10$^{22}$ ~H~cm$^{-2}$. Abs1 is proportional to the 6.4 keV-line flux, and hence is due to the MCs, while Abs2 is typical of interstellar absorption toward the Galactic center. Assuming that the GCPE plasma is spherically-extended around Sgr A* with a uniform density and the same angular distribution of the two temperature components, we quantitatively estimated the line-of-sight positions of the MCs from the flux ratio the GCPE spectrum suffered by Abs1 and that with no Abs1. The results suggest that the Sgr B MCs are located at the near side of Sgr A* in the GCPE.

\end{abstract}

\section{Introduction}
\begin{figure*}[!ht]
  \begin{center}
    \FigureFile(18cm,){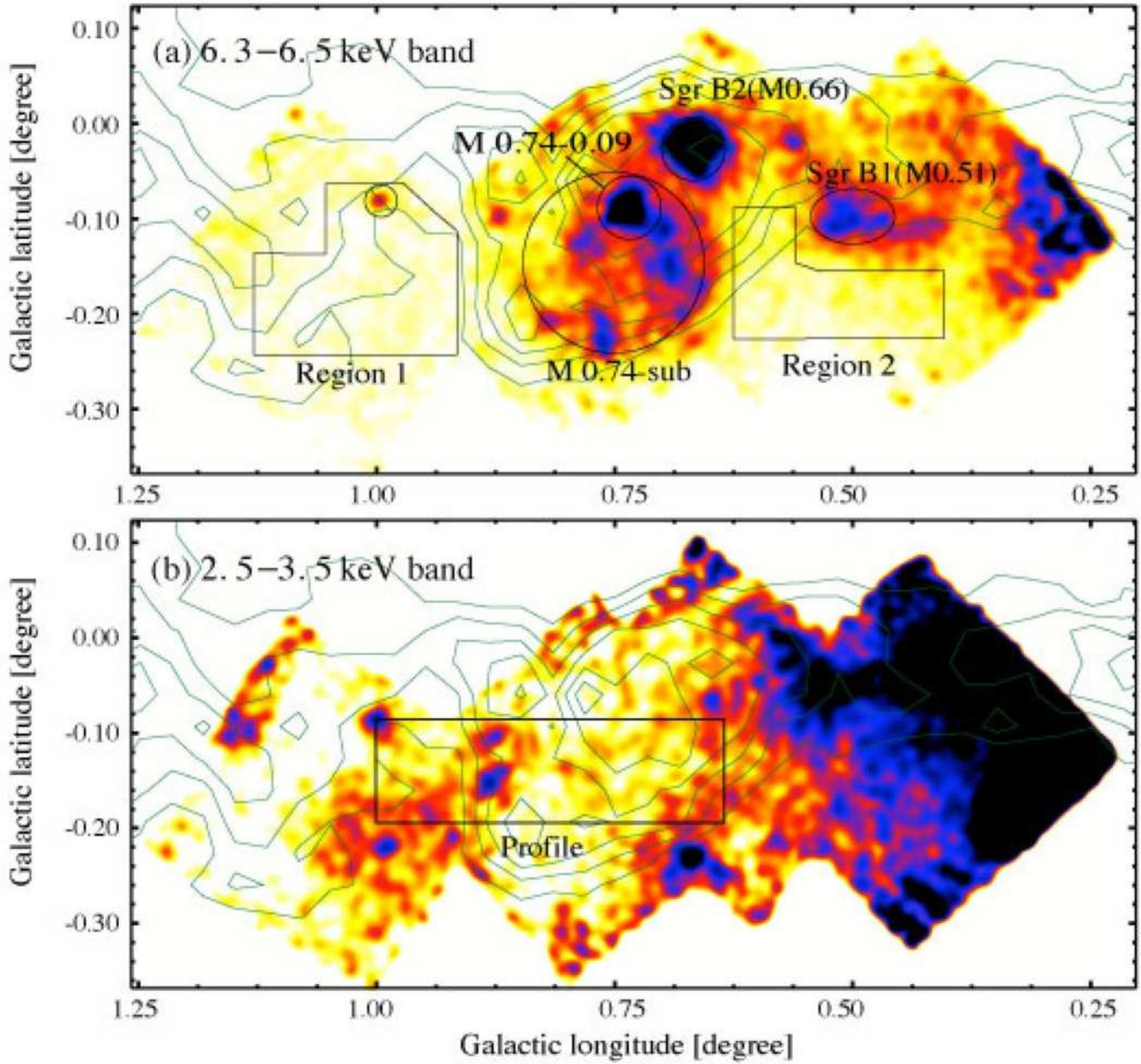}
  \end{center}
  \caption{XIS band images around the Sgr B region smoothed with a Gaussian kernel of $\sigma=\timeform{1'}$. (a) The 6.4 keV band (6.3--6.5 keV), which contains the Fe\emissiontype{I}~K$\alpha$ line, tracing the dense molecular clouds. (b) Medium band (2.5--3.5 keV) showing the distribution of the GCDX in low energy. The CS-line contours (\cite{tsuboi99}) in velocity range of $-$50--140 km s$^{-1}$ are added with green solid lines in (a) and (b). A hint of anti-correlation between (a) and (b) near $l\sim$\timeform{0.75D} would be due to strong absorption by the molecular clouds. Spectra regions are shown in black solid lines in (a); the spectrum region for M\,0.74-sub is the largest circle excluding the small circle of M\,0.74-0.09. The bright point source (SAX\,J1748.2$-$2808; \cite{nobu09}) is excluded for Region~1. The scan profile in figure \ref{profile} is made from the large rectangle (0.4 degree $\times$ 0.1 degree) in (b).
}\label{fig1}
\end{figure*}
The Galactic center (GC) region is complex with many molecular clouds (MCs), emission nebulae, supernova remnants (SNR), non-thermal radio emission, and so on. The major component is the central molecular zone (CMZ, $|l|\le\timeform{1D}$; \cite{moris96}) which contains mass of $3\times 10^{7} M_{\solar}$ (\cite{dahmen98}), approximately corresponds to 10\% of the whole molecular gas in the Galaxy. Sgr B1 and B2 are the most well-known and massive MCs in the CMZ. The detail of the CMZ structure has been studied mainly based on radio observations. Using ($l$,$V$) diagrams of the CO and CS lines and based on the assumption of the dynamical motions of the GC region, the line-of-sight distribution of MCs was investigated (e.g. \cite{sofue95}; \cite{tsuboi99}; \cite{nakanishi03}). On the other hand, \citet{sawada04} examined the spatial correlation between 2.6-mm CO line and  18-cm OH absorption line, and predicted the line-of-sight positions of MCs with no assumption of dynamical motions.
\par  
The GC region within the CMZ is also bright in the diffuse X-rays called the Galactic center diffuse X-ray (GCDX) (e.g. \cite{koyama07c}). The GCDX has been found to have thermal emission from hot plasmas, and hence we refer to the emission as the Galactic center plasma emission (GCPE). The X-ray spectrum of the GCPE exhibits strong K-shell emission lines from highly ionized atoms such as Fe\emissiontype{XXV} (6.7 keV) and S\emissiontype{XV} (2.45 keV). \citet{koyama07c} found that the line-intensity ratios of Fe\emissiontype{XXVI}~K$\alpha$ (6.97 keV)/Fe\emissiontype{XXV}~K$\alpha$ (6.7 keV) and Fe\emissiontype{XXV}~K$\beta$ (7.88 keV)/Fe\emissiontype{XXV}~K$\alpha$ (6.7 keV) are about 0.33 and 0.09, respectively. These values indicate that the plasma is in a collisional ionization equilibrium (CIE) at the temperature of k$T\simeq6.5$ keV. On the other hand, the K-shell emission from lighter elements, like the S\emissiontype{XV}~K$\alpha$ line (2.45 keV), is due to a lower temperature plasma of $\sim 1$ keV. The GCPE extends over the GC region of about 1 degree in longitude with a monotonic decrease as the distance from Sgr A* increases (\cite{koyama89}, \cite{yamauchi90}). 
\par  
The GCDX has another component, non-thermal emission with the Fe\emissiontype{I}~K$\alpha$ line of 6.4 keV from neutral irons (hereafter 6.4 keV-line). The diffuse 6.4 keV-line in the Sgr B region was firstly found by the ASCA satellite (\cite{koyama96}). The Suzaku satellite deeply observed the Sgr B region and found three bright 6.4 keV sources, Sgr B2 (M\,0.66$-$0.02), M\,0.74$-$0.09 (\cite{koyama07b}), and Sgr B1 (M\,0.51$-$0.10; \cite{nobu08}), at the locations of the MCs identified by the CO~line and the CS~line (\cite{sofue95}; \cite{tsuboi99}). The X-ray spectra of these sources have common features: a large equivalent width ($EW\geq$ 1~keV) for the 6.4 keV-line and a strong absorption ($N_{\rm H} \geq 10^{23}$~H~cm$^{-2}$) feature for the continuum below $\sim4$~keV. 
In addition, a time variability of the 6.4 keV-line has been discovered from Sgr B2 (M\,0.66$-$0.02; \cite{koyama08}). These facts suggest that the origin of the 6.4 keV-line and the related continuum emission is due to fluorescence and Thomson scattering taking place in molecular clouds irradiated by an external source. We refer to this component as the X-ray reflection nebula emission (XRNE). The most probable external X-ray source is a past outburst of Sgr A* (\cite{koyama96}; \cite{inui09}).
\par  
In this paper, we propose a new methodology to derive positions of the Sgr B molecular clouds along the line of sight: a correlation study of the flux of the 6.4 keV-line from the MCs and the line-of-sight absorption ($N_{\rm H}$) of the GCPE. 
    
\section{Observations and Data Reduction}

\begin{table*}[!ht]
  \caption{Suzaku Observations near the Sgr B Region}
  \label{tab1}

  \begin{center}
\begin{tabular}{lcccc}
\hline \hline             
  Target name &Obs.ID& Pointing direction\footnotemark[$*$]  & Obs.date   & Effective exposure  \\ 
 &  &$\alpha $ (J2000.0) \hspace{7.5mm}  $ \delta$ (J2000.0) &   (yyyy-mm-dd)     & (ks)  \\
 \hline
   GC Sgr\,B2&100037060&   \timeform{17h47m30.60s}  \hspace{3.5mm}  \timeform{-28D26'36.6''} & 2005-10-10 & 76.6   \\
   GC Number2&500005010&   \timeform{17h47m04.63s} \hspace{3.5mm}  \timeform{-28D37'46.2''} & 2006-03-27 & 88.4   \\
     GC Sgr\,B East&501039010& \timeform{17h48m04.87s} \hspace{3.5mm} \timeform{-28D21'06.5''} & 2007-03-03 & 96.4   \\
     Sgr\,D SNR&502020010&  \timeform{17h48m46.13s} \hspace{3.5mm} \timeform{-28D07'38.6''}  & 2007-09-06 & 139.1  \\
\hline

\multicolumn{5}{@{}l@{}}{\hbox to 0pt{\parbox{180mm}
{\footnotesize
       \par\noindent
\footnotemark[$*$]{The center of the XIS field of view. }
   }\hss}
}

 \end{tabular}
  \end{center}
\end{table*}
Deep multiple pointing observations toward the Sgr B region were performed with the X-ray Imaging Spectrometer (XIS) at the focal planes of the X-ray Telescope (XRT) onboard the Suzaku satellite from October 2005 to September 2007. The XIS system consists of three sets of front-illuminated (FI) CCD cameras (XIS\,0, 2, and 3) and one set of back-illuminated (BI) CCD camera (XIS~1); each CCD chip contains 1024$\times$1024 pixels (1 pixel = 24~$\mu$m$\times$24~$\mu$m) for the \timeform{18'}$\times$\timeform{18'} field of view. Two calibration sources of $^{55}$Fe are installed to illuminate two corners of each CCD for absolute gain tuning. The details of Suzaku, the XIS and the XRT are given in \citet{mitsuda07}, \citet{koyama07a}, and \citet{ser07}, respectively. The observations were made in the normal clocking mode with read-out cycle of 8~s. We screened the data by excluding the events observed below the low and day-Earth elevation angles at \timeform{5D} and \timeform{20D}, in addition to the removal of events during passage of the South Atlantic Anomaly. The total effective exposure time was about 400~ks. The observation log is shown in table \ref{tab1}.
\par
We performed data reduction and analysis using HEADAS software version 6.5.1 and XSPEC version 11.3.2. The calibration database was the version released\footnote{http://www.astro.isas.ac.jp/suzaku/caldb/} on 2008-06-02. We constructed the non--X-ray background (NXB) data from night-Earth observations\footnote{http://www.astro.isas.ac.jp/suzaku/analysis/xis/nte/}, and subtracted the NXB from the images and the spectra. Since the relative gains and response functions of the FI CCDs (XIS\,0, 2 and 3) are essentially the same, we made merged spectra of the FI CCDs \footnote{The XIS\,2 suddenly became unusable on 2006 November 9 possibly due to a micro-meteoroid impact on the CCD. Therefore, after this epoch, the merged FI-CCD spectra were those of XIS\,0 and 3.}. 

\section{Analysis and Results} 
\subsection{Overall Features of the X-Ray Images} 

\begin{figure}[!ht]
\begin{center}
    \FigureFile(8.5cm,){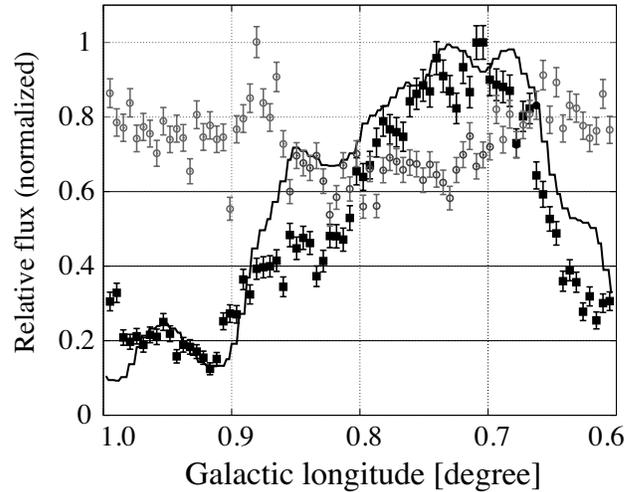}
\end{center}
  \caption{One-dimensional profiles of the Sgr B region along the Galactic longitude from $l=\timeform{1.0D}$--\timeform{0.6D} at $b=\timeform{-0.15D}\pm \timeform{0.05D}$ (see the solid rectangle in figure 1b).  Gray open circles and black filled squares are data of the 2.5--3.5 keV band and the 6.4 keV band (6.3--6.5 keV) with 1-$\sigma$ error bars, respectively. The radio CS~line profile ($-$50--140 km s$^{-1}$) is plotted with the black solid line. The flux of each data is normalized to its maximum value.}
\label{profile}
\end{figure}

Figure 1 shows XIS images of the Sgr B region in the (a) 6.3--6.5 keV and (b) 2.5--3.5 keV bands. The images were made by co-adding the FI and BI CCDs data, followed by correcting the exposure time and the vignetting effects of the XRTs. The data of the CCD corners illuminated (contaminated) by the calibration sources were excluded. To increase the visibility, we binned the images with 4$\times$4 pixels and smoothed them with a Gaussian kernel of $\sigma$ = $\timeform{1'}$.
\par  
The bright clumps in figure 1a are Sgr\,B2 (M\,0.66$-$0.02), M\,0.74$-$0.09, and Sgr\,B1 (M\,0.51$-$0.10). These are reported to be candidates of the X-ray reflection Nebula (XRN; \cite{koyama07b}; \cite{nobu08}). 
In addition, we found a new diffuse structure in the 6.4 keV-line at the south of M\,0.74$-$0.09 (hereafter, M\,0.74-sub; figure 1a). 
\par
The GCDX flux in the 2.5--3.5 keV band (figure 1b) is sensitive to absorption in $N_{\rm H}$ ranges of 10$^{22}$--10$^{24}$~H~cm$^{-2}$. 
This range corresponds to the interstellar absorption toward the GC region and the intra-cloud absorption of the MCs. 
Comparing figure 1a and 1b, we found a hint of anti-correlation between the 6.3--6.5 keV (6.4 keV-line) and the 2.5--3.5 keV band fluxes near $l=\timeform{0.75D}$: the 2.5--3.5 keV band flux is relatively weak in contrast with the strong emission of the 6.4 keV-line from the Sgr B MCs. 
\par
To confirm this anti-correlation, we made one-dimensional flux profiles in the 2.5--3.5 keV band and the 6.4 keV-line band along the Galactic longitude of $l=$\timeform{1.0D}--\timeform{0.6D} (see figure 2). For comparison, we also made profile and contours (in figure 1a and 1b) of the radio CS~line ($J=1$--0) by extracting the velocity-integrated flux in the range of $-$50--140 km s$^{-1}$ (\cite{tsuboi99}), within which most of the Sgr B MCs is included. The CS-line flux profile is added in figure 2. For all of the profiles, the vertical axis represents the sum-value at the Galactic latitude $b = \timeform{-0.15D}\pm\timeform{0.05D}$ (see the solid rectangle in figure 1b); the flux of each profile is normalized at the maximum value, and hence the relative flux range is from 0 to 1. 
\par
As shown in figure\,\ref{profile}, the flux profile of the 6.4 keV-line shows a bump at $l=$ \timeform{0.65D}--\timeform{0.85D}, which is in general agreement with that of the CS~line. The 2.5--3.5 keV band flux, on the other hand, shows an intensity drop at the bump of the 6.4 keV-line and the CS~line, which implies absorption due to the MCs. Since the 2.5--3.5 keV band flux is sum of the GCPE behind the MCs (should be anti-correlated to the 6.4 keV-line flux) and the GCPE in front of the MCs (no correlation to the 6.4 keV-line flux), the intensity drop would be smeared out to only $\sim15\%$. Detailed study of the spectra is given in section 3.2. We note that the angular size of the 6.4 keV-line bump is slightly narrower than that of the CS~line and that of the intensity drop in the 2.5--3.5 keV band at $l\sim \timeform{0.85D}$ (section 4.3).

\subsection{Overall Features of the Full-band Spectra} 

\subsubsection{Complex Components in the Spectra} 
\begin{figure*}[!ht]
  \begin{center}  
    \FigureFile(18.cm, ){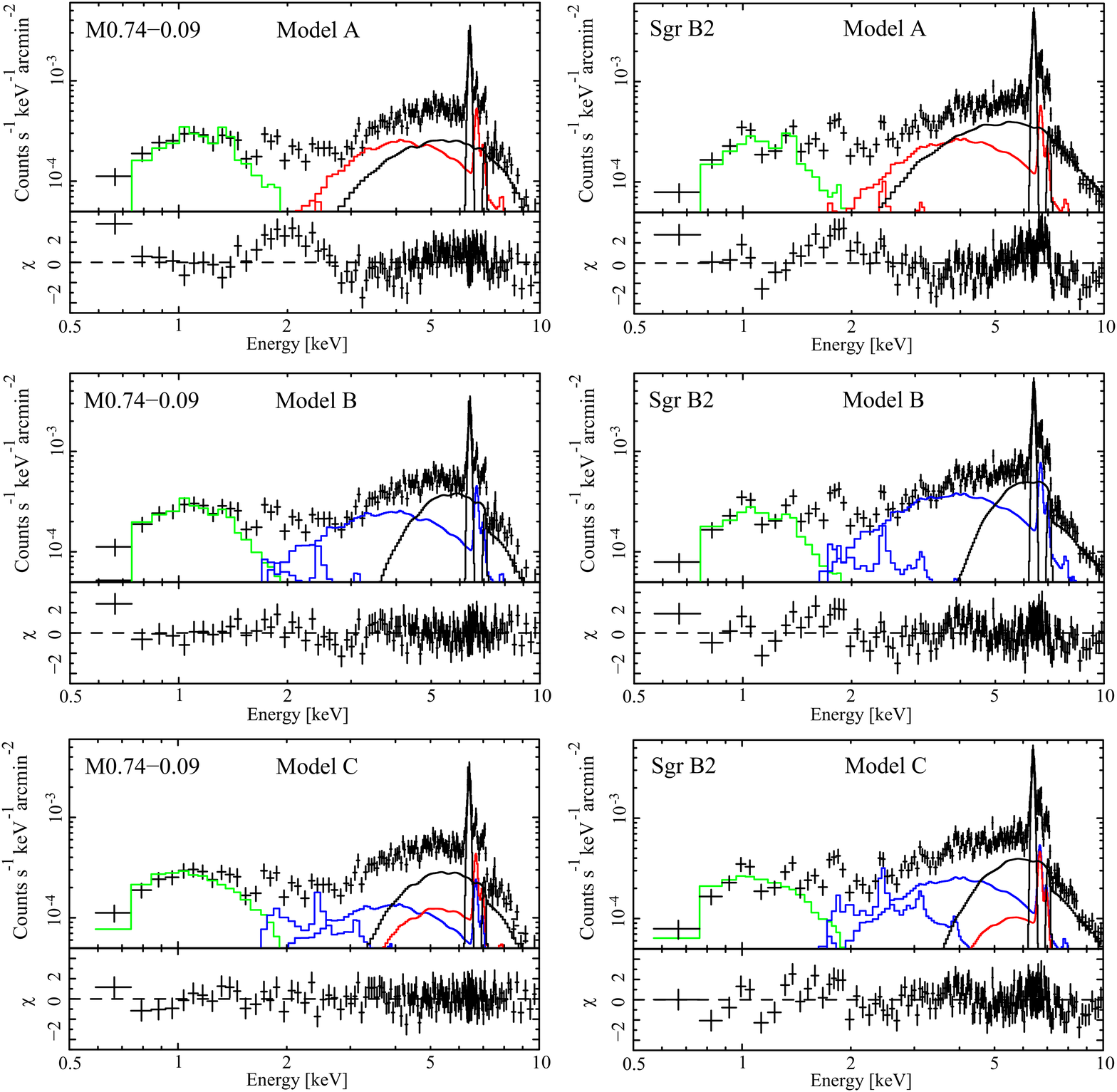}
  \end{center}
  \caption{FI spectra of M\,0.74-0.09 (left column) and Sgr\,B2 (right column) as examples for different models. The best-fit spectra and the residuals ($\chi$-distributions) of models A, B, and C are shown in the top, middle, and bottom rows, respectively. In each spectrum, components of the NGCE and the XRNE are shown by green and black solid lines; components of GCPE suffered by Abs1\,$\times$\,Abs2 and Abs2 are shown by the red and blue solid lines, respectively (see section 3.2.2).
}
\label{spec_mod}
\end{figure*}

We obtained X-ray spectra of the Sgr B region in the 0.5--10.0 keV band from the 6 solid line regions given in figure 1a. These are Sgr\,B2, M\,0.74$-$0.09, Sgr\,B1, M\,0.74-sub, Region\,1, and Region\,2. In figure\,\ref{spec_mod}, for brevity, we show 2 typical examples of the spectra.
\par
\par
The spectra have two notable common features (see figure\,\ref{spec_mod}). One is that the spectra exhibit many emission lines such as Fe\emissiontype{XXV}~K$\alpha$ (6.7 keV), Fe\emissiontype{XXVI}~K$\alpha$ (6.97 keV), Si\emissiontype{XIII}~K$\alpha$ (1.86 keV), S\emissiontype{XV}~K$\alpha$ (2.45 keV), and Fe\emissiontype{I}~K$\alpha$ (6.4 keV). These are K-shell lines from various atoms in various ionization states and hence the spectra are likely to consist of multi-temperature components, but may be reduced to at least three temperatures, which are responsible for the emission lines of the 6.4 keV (Fe\emissiontype{I}, cool gas), the 2.45 keV (S\emissiontype{XV}, hot plasma), and the 6.7 keV (Fe\emissiontype{XXV}, very hot plasma). 
The other feature is that the continuum shape shows broad local minima near 1--3 keV. Since the absorption of the GCDX spectrum is in the range of $6\times10^{22}$--10$^{24}$~H~cm$^{-2}$, the flux should rapidly drop below $\sim2$~keV. The presence of the local minima, therefore indicates that the full-band spectra have an additional soft X-ray component with no significant (small) absorption (the foreground emission). As a consequence, we decompose the spectra of the Sgr B regions to the following five components: 
\begin{description}
\item[(1)] Very hot plasma responsible for the 6.7 keV-line.
\item[(2)] Hot plasma responsible for the 2.45 keV-line.
\item[(3)] Cool gas component that emits the 6.4 keV-line. 
\item[(4)] Low energy (0.5--2 keV) X-rays with small absorption.
\item[(5)] The cosmic X-ray background (CXB).
\end{description}
\par
\citet{koyama07c} studied component~(1) using the iron/nickel line flux ratios of K~$\alpha$ and K~$\beta$ in the GC region. They concluded that component~(1) is a plasma of temperature k$T\simeq$ 6.5 keV in collisional ionization equilibrium. The iron abundance was determined to be $\simeq1$~solar\footnote{The solar abundance in this paper is referred to \citet{anders98}.}. We therefore use an APEC model\footnote{Astrophysical Plasma Emission Code: A model of an emission spectrum from collisionally-ionized diffuse gas. (\cite{smith01}).} for component~(1) and express it as APEC1 with fixed temperature and abundances of 6.5 keV and 1 solar, respectively. 
\par
With the Suzaku GC survey observations, many SNR candidates were found in the 2.45 keV-line band images. The plasma temperatures are nearly the same at about 1 keV. (\cite{nobu08}; \cite{mori08}; \cite{tsuru09}). The abundances are slightly different in each element, but are consistent with 1 solar. The abundances of S, the most prominent K$\alpha$ line emitter, are 0.6--1.3 solar (hereafter, the numbers in the parenthesis are in 90\% error range) for G\,0.42$-$0.04 (\cite{nobu08}), 1.2--1.5 solar for G\,359.79$-$0.26 (\cite{mori08}), and 1.1--2.7 solar for G\,359.41$-$0.12 (\cite{tsuru09}). Removing these SNR candidates, we still found the 2.45 keV line prevailing in all of the GC region (component~(2)).  Thus we use an APEC model for component~(2) (APEC2), assuming that the temperature and abundances to be $\sim1$ keV (free) and 1 solar (fixed), respectively, the same values to those of SNR candidates near the GC region.

\par
Component~(3) is emission possibly produced by the fluorescence and reflection from the MCs (XRNE): the fluorescent Fe\emissiontype{I}~K$\alpha$ (6.4 keV-line) with Thomson scattering continuum. The spectra of XRNs in the Sgr\,B and Sgr\,C complexes were extensively studied with Suzaku (\cite{koyama07b}; \cite{nobu08}; \cite{nakajima09}). Following these results, we use a phenomenological model: a power-law plus 2 Gaussian lines for the Fe\emissiontype{I}~K$\alpha$ and K$\beta$ as,
{\small
\begin{eqnarray}
\label{XRNE_model}
 {\rm XRNE}= A\times(E/{\rm keV})^{-\Gamma} + {\rm Gaussian1+ Gaussian2} \nonumber \hspace{5mm}\\  
{\rm [{\rm photons~cm}^{-2}{\rm~s}^{-1}{\rm~arcmin^{-2}}] }
\end{eqnarray}
}
 where the Gaussian center energy and relative intensity ratio of the Fe\emissiontype{I} K$\alpha$, K$\beta$ lines are fixed to 6.4 keV and 7.05 keV, and $1:0.125$, according to the theoretical values (\cite{kaa93}). 
\par
Component~(4) is low energy ($\leq2$~keV) emission extending to nearly 0.5~keV, indicating small or no absorption, and hence the foreground emission, which is unrelated to the GC region. The origin is unknown but would be either a local Galactic plasma (\cite{ebi08}) or unresolved faint dM stars (\cite{masui09}). Apart from the real origin, the spectra of component (4) contains faint line-like structure (see figure\,\ref{spec_mod}), we therefore used an APEC model (APEC3). Since the contribution of component~(4) above $\sim2$ keV is small, the fitting results of the spectra above $\sim2$ keV depend very weakly on the model assumption of component~(4).
\par
Component~(5) is the cosmic X-ray background (CXB). We applied the CXB model obtained from the Suzaku data at the north ecliptic pole,
{\small
\begin{eqnarray}
\label{CXB_model}
 {\rm CXB}=7.4\times10^{-7}\times(E/{\rm keV})^{-1.486} \nonumber \hspace{25mm} \\
 \hspace{15mm}{\rm [ photons~cm}^{-2}{\rm~s}^{-1}{\rm~arcmin^{-2}}]
\end{eqnarray}
}
As a matter of fact, the surface brightness of the CXB (see also \cite{kushino02}) in the Sgr\,B region is far lower than those of the other components by more than two orders of magnitudes, and hence the contribution of the CXB is almost negligible.
\vspace{3mm}
\subsubsection{Model Constructions and Fitting}
For the full-band spectra of the five components, we tried model fitting with many free parameters. However, due to large errors, no significant constraint on the best-fit parameters was obtained for any further scientific study and discussion. We therefore set several constraints on the model based on reasonable assumptions. 
\par
We assume that the spectral shape of the GCPE, which comprises the two plasmas (APEC1 and APEC2), is the same in all of the Sgr B regions. We thus combine components~(1) and (2), 
{\small
\begin{eqnarray}
\label{GCPEmodel}
 {\rm GCPE=APEC1}+{\rm APEC2}\hspace{35mm} \nonumber \\
\hspace{37mm}{\rm [photons~cm}^{-2}{\rm s}^{-1}{\rm arcmin^{-2}] }
\end{eqnarray}
} 
The temperature and abundances of APEC1 are fixed to 6.5 keV and 1 solar, respectively. We define a parameter $\alpha$, the normalization ratio of APEC1 and APEC2, assuming the common value in all regions near Sgr B (see figure \,\ref{longpro}). Thus, free parameters for the GCPE spectra are the temperature and the normalization factor of APEC2, and $\alpha$ with the same value among the relevant regions.
\par
The photon indices (${\Gamma}$) of the continuum of XRNs in Sgr B are scattered. This may be due to the limited energy band fittings in former studies (section 4.2). On the other hand, the equivalent width values of the 6.4 keV-line are nearly the same (\cite{koyama07b}, \cite{nobu08}, \cite{nakajima09}). Since the emission mechanisms would be the same for all Sgr B XRNs, we assume that the photon index (${\Gamma}$) and the equivalent width ($EW$) of the 6.4 keV-line are the same in all regions of Sgr~B. 
\par
\begin{figure}[!ht]
  \begin{center}
    \FigureFile(8.5cm, ){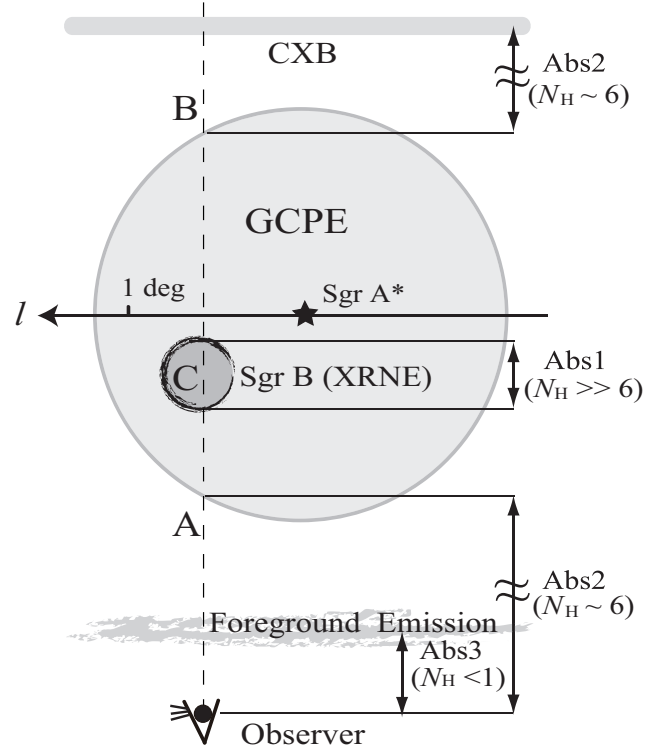}
  \end{center}
  \caption{Schematic view of the MCs distribution along the line of sight for Model C. The unit of absorption ($N_{\rm H}$) in the parentheses is 10$^{22}$~H~cm$^{-2}$. }
\label{abs}
\end{figure}

Although each of components~(1)--(5) should have a different absorption, we truncated them into three absorptions: Abs1, Abs2, and Abs3. Abs\,$i$ ($i=1,\,2,\,3$) is the integrated value along the line of X-ray path into our sight, which is given as $\exp(-N_{\rm H}\times\sigma_{(E)})$; $N_{\rm H}$ and $\sigma_{(E)}$ are respectively the hydrogen column density and absorption cross section with solar abundance. A schematic view of these three absorptions is given in figure\,\ref{abs}.
\par
Abs1 is the largest absorption ($N_{\rm H} \ge 10^{23}$~H~cm$^{-2}$), which takes place in/through the MCs. 
Abs2 is the interstellar absorption to/beyond the GC of $N_{\rm H}\sim6\times10^{22}$~H~cm$^{-2}$ (\cite{sakano02}). 
Abs3 is the smallest absorption applied to the foreground emission ($N_{\rm H} < 10^{22}$~H~cm$^{-2}$). 
\par
\par
In all 6 regions, we needed additional common components~(4) and (5), which are not due to the emission from the GC region. We combined these components and define them as the non GC emission (NGCE):
{ \small
\begin{eqnarray}
\label{NGCE_model}
 {\rm NGCE= Abs1\times Abs2 \times Abs2\times CXB +Abs3\times APEC3} \nonumber \\
\hspace{30mm}{\rm [photons~cm}^{-2}{\rm s}^{-1}{\rm arcmin^{-2}] }
\end{eqnarray}
}
The term Abs2$\times$Abs2$\times$CXB indicates that the CXB suffers from interstellar absorption (Abs2) two times: coming from the front and back sides of the GCPE. Here we assume that the absorption due to the back side on the GC is same as that on the front side of the GC.
 
\par
Referring the schematic view of the geometry of the MCs (XRNE) and the GCPE, we tried two extreme cases (see figure\,\ref{abs}).
(A): a MC is located at the near-side edge of the GCPE, then the XRNE and GCPE would have nearly the same absorption of Abs1$\times$Abs2. 
(B): a MC is located at the far-side edge of the GCPE; thus, the XRNE and the GCPE would have independent absorptions, given as Abs1$\times$Abs2 and Abs2. Model A and Model B, therefore, are given as,
{\small
\begin{eqnarray}
\label{AB}
{\rm (A): \hspace{3mm} Abs1\times Abs2 \times (XRNE + GCPE)+ NGCE} \hspace{8mm} \nonumber \\
{\rm (B): \hspace{3mm} Abs1\times Abs2\times  XRNE + Abs2\times GCPE +  NGCE}\nonumber \\
{ \rm [photons~cm^{-2}~s^{-1}~arcmin^{-2}] }  \hspace{0mm}
\end{eqnarray}
}
In the fittings, the free parameters are Abs1, Abs2, and the normalization factors of the 6.4keV-line flux, the GCPE, and the APEC3. While the mixing ratio $\alpha$ of APEC1 and APEC2 (in the GCPE), Abs3, the temperature and the abundance of APEC3, the equivalent width ($EW$) of the 6.4 keV-line and the photon index ($\Gamma$) for the power-law spectrum of XRNE are also free, but are assumed to be the same (common) in all relevant regions. We list the fitting parameters according to their properties in table\,\ref{para}.
\par  

\begin{table*}[ht]
  \caption{Summary of free and fixed parameters in the fittings for Model A, B, and C.}
   \label{para}
   \begin{center}
 \begin{tabular}{cccccc}
 \hline
\hline     
\multicolumn{6}{c}{Free parameters} \\
\hline
\multicolumn{1}{l}{Absorption:}&\multicolumn{5}{l}{$N_{\rm H}$ (Abs1), $N_{\rm H}$ (Abs2)}\\
\multicolumn{1}{l}{XRNE$^{*}$:}&\multicolumn{5}{l}{normalization (Fe\emissiontype{I} K$\alpha$ 6.4 keV line)}\\
\multicolumn{1}{l}{GCPE$^{*}$:}&\multicolumn{5}{l}{normalization (APEC2), position $R^{\dagger}$}\\
\multicolumn{1}{l}{NGCE$^{*}$:}&\multicolumn{5}{l}{normalization (APEC3)}\\

\hline
\multicolumn{6}{c}{Free parameters but common in all regions} \\
\hline
\multicolumn{1}{l}{Absorption:}&\multicolumn{5}{l}{$N_{\rm H}$ (Abs3)}\\
\multicolumn{1}{l}{XRNE$^{*}$:}&\multicolumn{5}{l}{photon index $\Gamma$, equivalent width $EW_{\rm 6.4 keV}$}\\
\multicolumn{1}{l}{GCPE$^{*}$:}&\multicolumn{5}{l}{k$T$ (APEC2), $\alpha$ (normalization ratio of APEC1/APEC2)}\\
\multicolumn{1}{l}{NGCE$^{*}$:}&\multicolumn{5}{l}{k$T$ (APEC3), abundances $Z$ (APEC3)}\\
\hline
\multicolumn{6}{c}{Fixed parameters} \\
\hline
\multicolumn{1}{l}{XRNE$^{*}$:}&\multicolumn{5}{l} {Fe\emissiontype{I} line energy K$\alpha$ = 6.4 keV, K$\beta$ = 7.05 keV, flux ratio (K$\beta$/K$\alpha$) = 0.125}\\
\multicolumn{1}{l}{GCPE$^{*}$:}&\multicolumn{5}{l}{k$T$ (APEC1) = 6.5 keV, $Z$ (APEC1 \& APEC2) = 1 solar}\\
\multicolumn{1}{l}{NGCE$^{*}$:}&\multicolumn{5}{l}{CXB = equation 2}\\

\hline
\multicolumn{6}{@{}l@{}}{\hbox to 0pt{\parbox{155mm}
{\footnotesize
       \par\noindent
       \footnotemark[$*$] See equation 1, equation 3, equation 4 for the definitions of XRNE, GCPE, and NGCE, respectively.
       \par\noindent
       \footnotemark[$\dagger$] For Model C only, see equation 6 and the text.

   }\hss}
}
 \end{tabular}
  \end{center}
\end{table*}


\begin{table*}[ht]
  \caption{Results of best-fit parameters$^{*}$ with Model A and B.}
   \label{tAB}
   \begin{center}
 \begin{tabular}{lcccccc}
 \hline
\hline     
  &\multicolumn{2}{c}{MODEL A}& &\multicolumn{3}{c}{MODEL B}\\
\hline
Region  &$N_{\rm H}^{\dagger}$ (Abs1$\times$Abs2)&6.4 keV-line$^{\ddagger}$&&$N_{\rm H}^{\dagger}$ (Abs1)&$N_{\rm H}^{\dagger}$ (Abs2)& 6.4 keV-line$^{\ddagger}$ \\
\hline
Sgr B2    &10.9$^{+0.9}_{-0.5}$ & 3.0$^{+0.1}_{-0.1}$&&54.8$^{+2.8}_{-2.4}$&5.8$^{+0.4}_{-0.1}$&6.6$^{+0.2}_{-0.2}$\\
M0.74-0.09&10.7$^{+0.5}_{-0.4}$&2.5$^{+0.1}_{-0.1}$&&40.4$^{+2.0}_{-1.7}$&5.2$^{+1.0}_{-0.8}$&4.3$^{+0.1}_{-0.6}$\\
Sgr B1    &8.4$^{+0.2}_{-0.2}$&1.53$^{+0.04}_{-0.08}$&&26.3$^{+2.0}_{-1.8}$&6.3$^{+0.2}_{-0.1}$&2.2$^{+0.1}_{-0.1}$\\
M0.74-sub &7.4$^{+0.1}_{-0.4}$&1.25$^{+0.02}_{-0.02}$&&21.9$^{+0.6}_{-1.0}$&5.0$^{+0.1}_{-0.1}$&1.63$^{+0.03}_{-0.02}$\\
Region1   &6.1$^{+0.1}_{-0.1}$&0.24$^{+0.01}_{-0.01}$ &&18.6$^{+1.9}_{-1.2}$&4.6$^{+0.1}_{-0.1}$&0.33$^{+0.01}_{-0.01}$\\
Region2   &7.7$^{+0.1}_{-0.3}$&0.43$^{+0.01}_{-0.03}$ &&31.8$^{+2.2}_{-1.9}$&5.8$^{+0.1}_{-0.2}$&0.76$^{+0.02}_{-0.04}$\\
\hline
 \multicolumn{1}{l}{Region-common parameters}&\multicolumn{5}{l}{}\\
\hline 
k$T$ (keV) (APEC2)&\multicolumn{2}{c}{0.70 (0.68--0.72)}&& \multicolumn{3}{c}{0.69 (0.68--0.74)}\\
$\alpha$ ($\frac{\rm APEC1\,norm}{\rm APEC2\,norm}$) &\multicolumn{2}{c}{0.47 (0.46--0.48)}&& \multicolumn{3}{c}{ 0.59 (0.57--0.60)}\\
Photon index $\Gamma$  &\multicolumn{2}{c}{0.52 (0.47--0.59) }&& \multicolumn{3}{c}{2.16 (2.10--2.23)}\\
$EW_{\rm 6.4 keV}$ (keV) &\multicolumn{2}{c}{ 1.78 (1.75--1.80)}&& \multicolumn{3}{c}{1.31 (1.28--1.34)}\\
Abs3 ($10^{22}$\,H\,cm$^{-2}$) &\multicolumn{2}{c}{ 0.21 (0.21--0.22)}&& \multicolumn{3}{c}{0.20 (0.19--0.21)}\\
k$T$ (keV) (APEC3)&\multicolumn{2}{c}{1.02 (0.99--1.05)}&& \multicolumn{3}{c}{0.93 (0.92--0.94)}\\
$Z$ (solar) (APEC3)&\multicolumn{2}{c}{0.004 (0.002--0.006)}&& \multicolumn{3}{c}{0.008 (0.007--0.011) }\\
\hline

$\chi^{2}$/d.o.f\,$^{\S}$ &\multicolumn{2}{c}{$4749/2537=1.872$}&& \multicolumn{3}{c}{$4479/2537=1.765$} \\
\hline
\multicolumn{6}{@{}l@{}}{\hbox to 0pt{\parbox{135mm}
{\footnotesize
       \par\noindent
       \footnotemark[$*$]The uncertainties are at 90\% confidence level.
       \par\noindent
       \footnotemark[$\dagger$]The value of $N_{\rm H}$ in the unit of $10^{22}$\,H\,cm$^{-2}$.
       \par\noindent
       \footnotemark[$\ddagger$]Flux in the unit of $10^{-6}$photons cm$^{-2}$s$^{-1}$arcmin$^{-2}$.
        \par\noindent
       \footnotemark[$\S$]The results obtained by the simultaneous fitting of the 12 spectra (the FI and BI spectra of 6 regions).
   }\hss}
}
 \end{tabular}
  \end{center}
\end{table*}
We performed a simultaneous fitting for the $2\times6$ spectra (the FI and BI spectra of 6 regions) with Model A and Model B. The best-fit parameters and $\chi^2$/d.o.f are listed in table\,\ref{tAB}. As a results, both models were rejected by the $\chi^2$/d.o.f test. Model~A gave a large $\chi^2$/d.o.f of 1.87, and shows large residuals in the medium range of 1--3 keV and the hard range of 5--7 keV (see figure \ref{spec_mod}). The best-fit values of Abs1\,$\times$\,Abs2 (with Model~A) is $\simeq10\times10^{22}$~H~cm$^{-2}$ (see table\,\ref{tAB}), which are close to the typical value toward the GC region. These values may be reasonable for the GCPE but could not be applied to the XRNE, because the absorption of each XRNE was found to be $>10^{23}$~H~cm$^{-2}$ by independent analysis of XRN (\cite{koyama07b}; \cite{nobu08}; \cite{nakajima09}). 
\par    
In Model~B, the apparent residuals in the medium and hard range found in Model~A disappeared partially, and hence $\chi^2$/d.o.f was improved to 1.77. We still see unnatural residuals between 2--4 keV. Although the $\chi^2$/d.o.f of 1.77 is large enough to reject Model~B at more than the 99.9\% confidence level, this value is slightly smaller than that of Model~A. At the end of this section, we discuss why we reject Model~B from a different point of view.
\par    
We then constructed Model C, adding a new free parameter $R$ (for each region), in which a MC/XRN is inside the GCPE with a relative position of $R$ (0--1); the smaller $R$ indicates the nearer (front) side.    
{\small            
\begin{eqnarray}
\label{C}
{ \rm (C): \hspace{2mm} Abs1\times Abs2 \times(XRNE} + (1-R)\times {\rm GCPE)}\hspace{5mm} \nonumber\\
{\rm  +Abs2\times} (R\times {\rm GCPE) +NGCE} \hspace{20mm} \nonumber\\
 \hspace{35mm} {\rm [photons~cm^{-2}~s^{-1}~arcmin^{-2}] } 
\end{eqnarray}
}
We performed a simultaneous fitting for Model~C by the same procedure as for Model~A and B. Although Model C has six new free parameters ($R$ for the six regions) more than Model A and Model B, the degree of freedom (d.o.f) of Model C (2531) is almost equal to that of Model A and and B (2537). On the other hand, the $\chi^2$/d.o.f of Model C is largely improved to 1.34 from those of Model A (1.87) and Model B (1.77).
\par
One may argue that the spectra are contaminated by unresolved point sources in the Sgr~B region. In fact, \cite{koyama09} estimated that the point source fluxes provide a significant fraction of the GCDX (GCPE + XRNE) emission, up to about  1/6 near Sgr A*, which may partially contribute to the 6.7 keV-line and the 6.4 keV-line fluxes in the GCPE and XRNE, respectively. According to the Chandra cataloged sources reported by \citet{muno06}, the number of point sources per solid angle near Sgr B2 is $\sim 10\%$ compared to that near Sgr~A*, while \citet{nobu08} shows that the GCDX flux near Sgr~B2 is also $\sim10\%$ of that near Sgr~A*. We can infer that the contribution of point sources to GCDX would be roughly the same in the Sgr B region. The point source contribution may cause possible systematic errors in the present spectral analysis. The most serious errors, therefore, should be the region-to-region fluctuation of the fractions of the point source fluxes, which may be rather small effects in a relative comparison of the best-fit parameters. At this moment, we have no information on the exact flux and spectrum of the integrated point sources in the Sgr~B region. We therefore ignore the point source contribution. 
\par
All together, taking account of possible systematic errors due to the point source contribution and those due to the simplified model, we regard Model~C as a good approximation of the overall spectra in the Sgr B region, although the $\chi^2$/d.o.f of 1.34 is statistically marginal to be acceptable. The best-fit common parameters for all regions are as follows (hereafter, the numbers in the parenthesis are in 90\% error range). The temperature of APEC2 is 0.87 (0.81--0.91) keV and the mixing ratio ($\alpha$) in the GCPE is determined to be 0.27 (0.26--0.28). The photon index ($\Gamma$) and  the equivalent width ($EW$) of the 6.4 keV-line for the XRNE are 1.72 (1.64--1.80) and 1.59 (1.54--1.63) keV, respectively. The absorption (Abs3), temperature (k$T$) and the abundance of the foreground emission (APEC3) are 0.17 (0.16--0.18)$\times10^{22}$ H cm$^{-2}$, 0.85 (0.83--0.87) keV, and 0.011 (0.009--0.014) of solar, respectively. The other best-fit parameters of Model~C for each region are listed in table\,\ref{C}.
\par

\begin{table*}[ht]
  \caption{Detailed results of best-fit parameters with Model C.}
   \label{C}
   \begin{center}
 \begin{tabular}{lcccccc}
 \hline
\hline     
Region&$N_{\rm H}^\dagger$ (Abs1)&$N_{\rm H}^\dagger$ (Abs2)&6.4 keV-line$^{\ddagger}$& APEC3 norm$^\S$ &GCPE norm$^\|$&Position $R^\#$ \\
\hline
Sgr B2    &36.5$^{+2.4}_{-2.2}$ &8.2$^{+0.3}_{-0.5}$ &5.1$^{+0.3}_{-0.3}$   & 0.09$^{+0.01}_{-0.01}$&3.40$^{+0.36}_{-0.48}$&0.44$^{+0.06}_{-0.06}$ \\ 
M0.74-0.09&22.9$^{+2.0}_{-1.8}$ &6.8$^{+0.4}_{-0.4}$ &3.3$^{+0.2}_{-0.2}$   & 0.11$^{+0.01 }_{-0.01}$&2.67$^{+0.29}_{-0.28}$&0.31$^{+0.04}_{-0.04}$\\ 
Sgr B1    &12.3$^{+1.0}_{-1.0}$ &7.0$^{+0.2}_{-0.3}$ &1.8$^{+0.1}_{-0.1}$   & 0.11$^{+0.01}_{-0.01}$ &3.46$^{+0.25}_{-0.38}$&0.41$^{+0.06}_{-0.06}$ \\ 
M0.74-sub &11.5$^{+0.5}_{-0.5}$ &5.3$^{+0.1}_{-0.2}$ &1.42$^{+0.03}_{-0.03}$ & 0.090$^{+0.003}_{-0.004}$&1.66$^{+0.10}_{-0.09}$&0.26$^{+0.04}_{-0.04}$ \\ 
Region1   &7.2$^{+0.2}_{-0.2}$  &3.6$^{+0.2}_{-0.3}$ &0.26$^{+0.01}_{-0.01}$ & 0.105$^{+0.006}_{-0.004}$ &1.40$^{+0.05}_{-0.04}$&0.15$^{+0.01}_{-0.01}$\\ 
Region2   &12.0$^{+0.6}_{-0.8}$  &6.3$^{+0.1}_{-0.1}$ &0.47$^{+0.03}_{-0.03}$ & 0.130$^{+0.004}_{-0.036}$ &2.88$^{+0.11}_{-0.18}$&0.34$^{+0.04}_{-0.04}$\\
\hline
$\chi^{2}$/d.o.f\,$^{**}$ &\multicolumn{6}{c}{$3382/2531=1.336$}\\
\hline
\multicolumn{6}{@{}l@{}}{\hbox to 0pt{\parbox{145mm}
{\footnotesize
       \par\noindent
       \footnotemark[$*$] The uncertainties are at 90\% confidence level.
       \par\noindent
       \footnotemark[$\dagger$] The value of $N_{\rm H}$ in the unit of $10^{22}$\,H\,cm$^{-2}$.
       \par\noindent
       \footnotemark[$\ddagger$] Flux in the unit of $10^{-6}$photons cm$^{-2}$s$^{-1}$arcmin$^{-2}$.
        \par\noindent
       \footnotemark[$\S$] The normalization factor for the foreground emission (c.f APEC3 in equation 4) of the APEC model re-normalized with the region size $A$ [arcmin$^2$]; it is expressed as $7.07 \times 10^{-11}/(4\pi D^2\,A)\,EM$ [cm$^{-5}$arcmin$^{-2}$], where $D$ and $EM$ are the distance to the source [cm], and the emission measure [cm$^{-3}$], respectively. 
       \par\noindent
  \footnotemark[$\|$] In same expression as $\S$, but the normalization factor of APEC2 in the GCPE (c.f equation 3 and text).  
       \par\noindent
       \footnotemark[$\#$] $R$ is the fraction of the GCPE suffered by Abs2, which indicates the positions of MCs (see equation 6 and figure\,\ref{pos}). 
       \par\noindent
       \footnotemark[${**}$] The results obtained by the simultaneous fitting of the 12 spectra (the FI and BI spectra of 6 regions).
   }\hss}
}
 \end{tabular}
  \end{center}
\end{table*}
  
\par
The 6.4 keV-line are likely to be fluorescence in the MC generated by an external X-ray source. Then the flux should be proportional to the product of the external X-ray flux and the number of iron atoms along the line-of-sight X-ray path in the MC, in the optically thin case at the energy near 6.4--7.1\footnote{E = 7.1 keV is the K absorption-edge of neutral iron.} keV ($N_{\rm H}\leq10^{24}$~H~cm$^{-2}$). The number of iron atoms is approximately proportional to the column density $N_{\rm H}$ of the MC. Therefore, in the case that the external X-ray fluxes are the same among the MCs, Abs1 should be approximately proportional to the 6.4 keV-line flux. 
\par
\begin{figure}[!ht]
  \begin{center}
    \FigureFile(8.5cm,){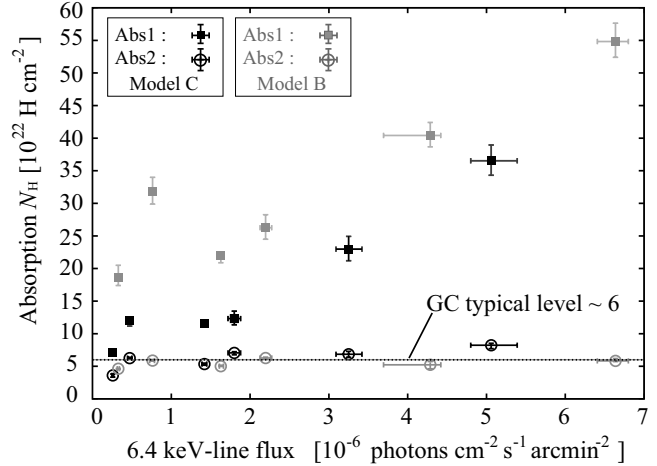}
  \end{center}
  \caption{Correlations between the 6.4keV-line flux and absorptions (made from data listed in table\,\ref{tAB} and table\,\ref{C}). 
Results of Model C and Model B are plotted by the black and gray marks; filled squares and open circles represent Abs1 and Abs2, respectively.}
\label{64_nh}
\end{figure}
In figure\,\ref{64_nh}, we plot the best-fit Model C values of Abs1 and Abs2, as a function of the 6.4 keV-line flux. For a comparison, we also plot the results of Model B. A notable fact to support Model C is that Abs1 is in good proportionality to the 6.4 keV-flux. This feature is in contrast to the results of Model B, where the proportionality is not very good in the low-flux range. Most importantly, Model B requires an extra off-set of about $2\times10^{23}$~H~cm$^{-2}$, which is un-realistic. This leads us to exclude Model B in addition to the  $\chi^2$/d.o.f test. We estimated $N_{\rm H}$ values of the Sgr B MCs from the data of the CO~line (\cite{sofue95}; \cite{dahmen98}) and the CS~line (\cite{tsuboi99}). They are in the range of $N_{\rm H}=10^{23}$--$10^{24}$~H~cm$^{-2}$, in good agreement with the Model~C prediction.
\par
As shown in figure\,\ref{64_nh} for Model~C, we found that Abs2 is almost constant at $N_{\rm H}\sim6\times10^{22}$~H~cm$^{-2}$. This is consistent with the view that Abs2 is interstellar absorption (see figure\,\ref{abs}) toward the GC of about $N_{\rm H}\simeq6\times10^{22}$~H~cm$^{-2}$ (\cite{reike89}; \cite{sakano02}).
\par

\section{Discussions}
\subsection{Justifications of Model Constraints \& Assumptions}
\begin{figure}[!ht]
  \begin{center}
    \FigureFile(8.5cm,){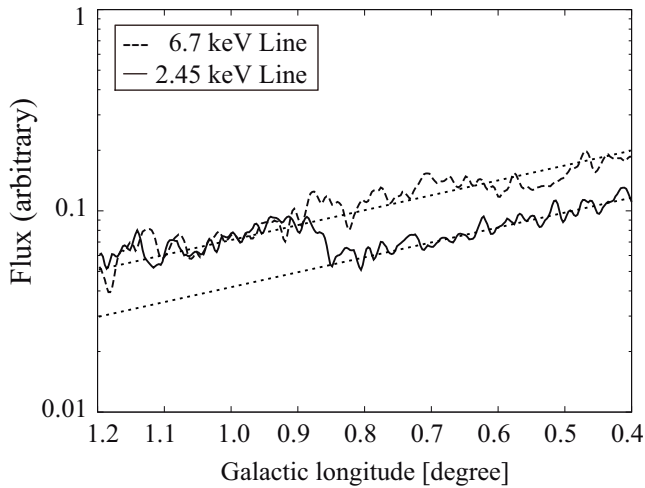}
  \end{center}
  \caption{Flux distribution of the 6.7 keV line (6.5 keV plasma; dashed line) and the 2.45 keV line ($\sim 1$ keV plasma; solid line) along the galactic longitude from $l = \timeform{1.2D}$--$\timeform{0.4D}$ at $b = \timeform{-0.15D}\pm\timeform{0.05D}$. Two parallel dotted lines guide eyes to see the same distribution of the 6.5 keV plasma and the $\sim 1$ keV plasma. }
\label{longpro}
\end{figure}

In the previous section, we introduced a complicated model to described the spectra of the GCDX.  In order to obtain meaningful values (no large errors) of physical parameters, we further added some constraints and assumptions. If, for example, we let the normalization ratio $\alpha$ (APEC1/APEC2) to be free for all regions, we can not give a significant constraint on the absorptions of GCPE and/or the temperature of APEC1, because the two plasma emission become nearly the same in the band near 2 keV. We summarize the adopted constraints and assumptions below for further discussions:

\begin{itemize}
\item The 6.5 keV plasma and $\sim 1$ keV plasma, which comprise the GCPE, 
have nearly the same angular distribution near the Sgr B region. (see figure\,\ref{longpro})
\item The abundances of the GCPE are nearly 1 solar, the same value as those of nearby new SNR candidates.  
\item The XRNE has a common photon index ($\Gamma$) and $EW_{\rm 6.4 keV}$ near the Sgr B region.
\item The point sources contribution to the GCDX is small and hence could be ignored in the Sgr B region.
\end{itemize}

The most fundamental constraint is the first item; the factor $\alpha$ is constant from region to region. This implies that the angular distributions of the very hot (6.5 keV) and hot ($\sim$1 keV) plasmas are nearly the same near Sgr B. In order to judge this constraint, we show the flux distribution of the 6.7 keV and 2.45 keV lines along the galactic longitude of $l = \timeform{1.2D}$--$\timeform{0.4D}$ at $b = \timeform{-0.15D}\pm\timeform{0.05D}$ (figure\,\ref{longpro}). The 6.7 keV and 2.45 keV lines are strong characteristic lines that well represent, respectively, the distributions of the 6.5 keV plasma and the $\sim1$ keV plasma with the least systematic errors.     

As is found in figure\,\ref{longpro}, the 2.45 keV line and the 6.7 keV line have almost the same slope (angular distribution). The small drop of the 2.45 keV-line flux near at $l = \timeform{0.85D}$--$\timeform{0.6D}$ would be due to absorption by the M\,0.74$-$0.09 MC and the drop at $l = \timeform{0.6D}$--$\timeform{0.4D}$ is due to the absorption by the Sgr B1 MC. This figure, therefore, indicates that the emission of the $\sim$1 keV plasma (APEC2) and the 6.5 keV plasma (APEC1) have a common normalization ratio ($\alpha$) for the GCPE near the Sgr B regions. The justification of the other constraints and assumptions listed in the last three items are separately discussed in the previous sections. We thus conclude that Model~C with the above listed constrains and assumptions provides a practical approximation for the overall spectra in the Sgr B region, although the $\chi^2$/d.o.f of 1.34 is statistically marginal to be acceptable.

\subsection{Line-of-Sight Positions of MCs}

\begin{figure}[!ht]
  \begin{center}
   \FigureFile(8.5cm,){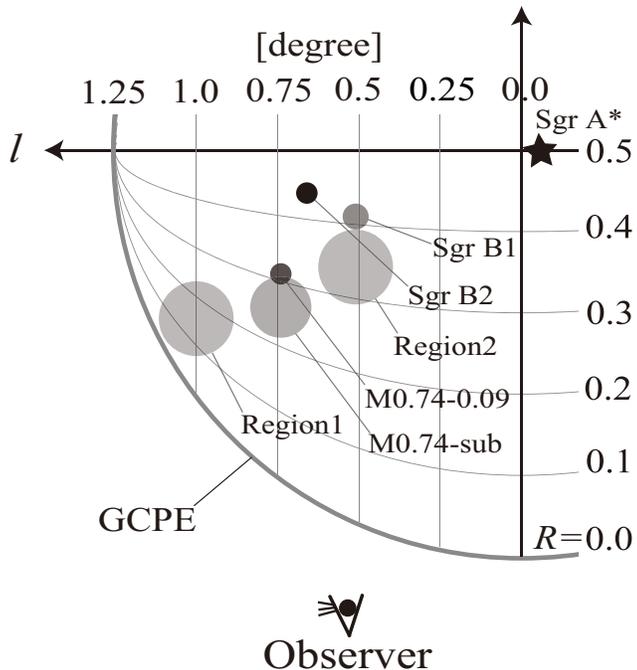}
  \end{center}
  \caption{Relative positions of each MC (filled circles) along the line of sight, where the horizontal lines of $R=0.0$ and 0.5 imply the anterior border line  and the GC line (Sgr A* is on this line), respectively. The circles filled with darker colors indicate MCs of the larger $N_{\rm H}$ values.}
\label{pos}
\end{figure}

From the results of section 3, we conclude that the MCs of Sgr B should be located in the GCPE (Model~C), neither in the near-side edge (Model~A) nor in the far-side edge (Model~B) of the GCPE. For simplicity, we assume that the volume emissivity of the GCPE is approximately a uniform sphere with a uniform plasma density on the Galactic plane. \citet{koyama89} and \citet{yamauchi90} reported that the 6.7 keV line distributed with a Gaussian shape of 1.8 deg (FWHM) (in parallel to the Galactic plane). We therefore take the boundary (radius) at $|l|=$\,1.25 deg, within which 90\% of the GCPE is included. The center of the GCPE is approximately Sgr A* (at $l$ = \timeform{-0.056D}; \cite{yusef99}). Then the parameter $R$ gives the line-of-sight positions of the MCs within the GCPE\footnote{The emission measure $EM$ can be written as $\int n_{\rm e}n_{\rm H}{\rm d}V = n_{\rm e}n_{\rm H}S\int{\rm d}l$ $\propto n_{\rm e}n_{\rm H}S L$, where $L$ is the length of the GCPE in the line of the sight, $S$ and $n_{\rm e}n_{\rm H}$ are the projected area and the product of densities of electrons and hydrogens, respectively. Then the parameter $R$ is the fraction of $EM^{\rm front}_{\rm GCPE}$/$EM^{\rm all}_{\rm GCPE}$, which is equal to $L^{\rm front}_{\rm GCPE}/L^{\rm all}_{\rm GCPE}$: the relative line-of-sight position of MCs in the GCPE.}, where $R=0$ and 0.5 indicate the anterior border line and the center line (Sgr A* is on this line) of the GCPE, repectively. We plot the best-fit position parameter ($R$) for the 6 regions (indicated in figure~1a) in the schematic view of figure\,\ref{pos}.
\par

As demonstrated in figure\,\ref{pos}, the MCs in the Sgr~B region are generally on the near-side with respect to Sgr A*. Also if all of the MCs/6.4 keV clumps, Sgr B1, Sgr B2, M0.74$-$0.09, and M0.74-sub are the constituent parts of the CMZ, the results indicate that the CMZ at the positive Galactic longitude has a bar-like distribution that generally inclines toward us. Although no further quantitative comparison is available with the present data, the inclining trend of the Sgr B MCs toward us is consistent with the radio molecular observations results based on the ($l,V$) diagrams study by \citet{sofue95} and the spatial correlation study between 2.6-mm CO line and 18-cm OH absorption line by \citet{sawada04}.
\par
It may be suspicious that the best-fit values of $R$ in Model~C indicate that the MCs are all closer to the near-side edge of the GCPE rather than to the far-side edge, while $\chi^2$/d.o.f. $=1.872$ of Model~A (MCs at the near-side edge) is rather worse than $\chi^2$/d.o.f. $=1.765$ of Model~B (MCs at far-side edge). As shown in equation~5, Model B have the extra-part Abs2$\times$GCPE which Model A does not have. This gives a slightly better $\chi^2$/d.o.f. of Model~B (1.765) than Model A (1.872). On the other hand, the improvement of Model C ($\chi^2$/d.o.f.$=1.336$) from Models A (and B) is far larger than that between Model~A and B.

\subsection{A Unified Picture of the XRN}
 
According to former studies (\cite{koyama07b}; \cite{nobu08}), the equivalent width of the 6.4 keV-line and the continuum photon index ($EW$, $\Gamma$) of the Sgr B XRNs have been reported to be (1.1~keV, 3.2) for Sgr B2, (1.55~keV, 1.4) for M\,0.74-0.09, and (1.4~keV, 1.8) for Sgr B1. The $\Gamma$ values are significantly scattered among the XRNs, but the variation of $EW$ is not very large. The GCPE flux is position-dependent, which means that a proper background-selection is difficult. The large scatter of $\Gamma$ values in the former studies may be due to the nearby GCPE-subtraction and would also be due to the band limited (e.g. 5--8 keV) spectral fitting. In this work (section 3.2 with Model C), the $EW$ and $\Gamma$ values are estimated by an independent method, a simultaneous fitting of a wide energy band (0.5--10.0 keV), in which nearby GCPE was not subtracted but directly included as the GCPE model. We also note that these uncertainties do not largely affect the $EW$ values. In fact, We found that the averaged $EW\simeq1.6$~keV, which is consistent with the previous studies of the Sgr B XRNs. \citet{koyama09} studied the GCDX near the Sgr A region ($|l|\le\timeform{0.2D}$), and found that the $\Gamma$ of the continuum flux is 1.9; this reduced to 1.4 if we exclude the point sources contribution. \citet{nakajima09} studied the 6.4 keV clumps in the Sgr C region and found $\Gamma$ to be in range of 1.6--1.9. We determined the common $\Gamma$ in Sgr B to be $\simeq1.7$; this value agrees with the results in Sgr A, B, and C. Thus, we infer that the XRNE may have a unified photon index of 1.4--1.9 in the overall GC region.  

\par     
\begin{figure}[!ht]
  \begin{center}
   \FigureFile(8.5cm,){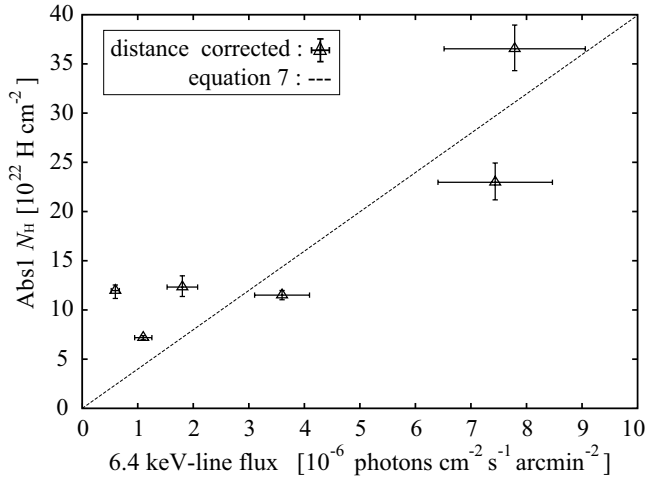}
  \end{center}
  \caption{The best-fit absorption (Abs1) vs the distance-corrected 6.4 keV-line flux from results of Model C in table\,\ref{C} (open triangles), where the correction factors are estimated from the distances between MCs and Sgr A* with figure\,\ref{pos}, and are normalized to that of Sgr B1.}
\label{64_cor}
\end{figure}

As discussed by \citet{koyama08} and \citet{nobu08}, a possible external X-ray source is Sgr A*, therefore the X-ray flux at the Sgr B MCs should be proportional to $L_{0}D^{-2}$, where $L_{0}$ and $D$ are, respectively, the luminosity of Sgr A* and the distance between Sgr A* and MCs. Thus the flux of the fluorescent 6.4 keV-line should be modified by this distance effect. In figure\,\ref{64_cor}, we plotted distance-corrected 6.4-keV fluxes as a function of Abs1, where the distances are estimated using figure\,\ref{pos}. We then fitted the correlation between the best-fit absorption (Abs1) and the 6.4 keV-line flux (distance-corrected), and obtained, 
\\
\begin{eqnarray}
N_{\rm H}\,({\rm Abs1}) = 3.99\, (\pm 0.56)\times I_{\rm 6.4 keV} \,\,\,\, [ 10^{22} \,\, {\rm H~cm}^{-2}] , \nonumber \\ 
  I_{\rm 6.4 keV \, }:[{\rm 10^{-6}~photons~s^{-1}~cm^{-2}~arcmin^{-2}}]
\end{eqnarray}
This relation is generally in agreement with the fluorescent process due to irradiation by Sgr A* with the past luminosity of 2--3$\times10^{39}$~ergs~s$^{-1}$. If $L_0$ was time constant, the relation between the distance-corrected 6.4 keV-line flux vs. Abs1 should be in good proportionality. However, we see some data scatters from the proportionality relation (figure\,\ref{64_cor}). It can be explained by assuming a time variable flux of the irradiating source Sgr~A*  (\cite{koyama08}; \cite{inui09}).
\subsection{The Correlation of CS~line and 6.4 keV-line Flux}
As shown in figure \ref{profile}, the 2.5--3.5 keV band flux shows an intensity drop at the bump of the 6.4 keV-line and the CS~line. The anti-correlation to the  2.5--3.5 keV band flux is clearer in the CS~line than that in the 6.4 keV-line. This may be because that the CS~line traces all (GC + foreground) MCs in the line of sight from which the GCPE may suffer absorption, while 6.4 keV-line only traces MCs near the irradiating source in the GC region. The better anti-correlation of the CS~line against the 2.5--3.5 keV band flux indicates that a large amount of MCs, identified by the velocity-integrated CS lines but not bright in 6.4 keV-line, is located in the near side of the GCPE. The angular size of the 6.4keV bump is slightly narrower than that of the CS~line ($l\sim\timeform{0.85D}$ and $\timeform{0.63D}$). The narrower angular size of the 6.4 keV bump than that of the CS~line may also be explained if Sgr A* is the external X-ray source to produce the 6.4 keV clumps. The surface brightness of fluorescent X-rays is faint at the near and far sides of the cloud-limb because of small $N_{\rm H}$ values, but becomes brighter when the external X-rays go into the large $N_{\rm H}$ ($\geq 10^{23}$~H~cm$^{-2}$) region in the MCs. Another possible reason of the faint 6.4 keV-line flux at the far-side cloud limb from Sgr A* is due to the time variability of the Sgr A* X-rays; an X-ray front of possible X-ray outbursts about 300 years ago (\cite{koyama08}; \cite{inui09}) does not yet reach at the far-side limb of $l\sim\timeform{0.85D}$. 
\\
\par
We are grateful to all members of the Suzaku team, especially H.~Matsumoto and H.~Uchiyama for their useful comments and supports. We also thank Y.~Hyodo and M.~Sawada for improving the draft quality. M.N. acknowledges financial supports from the Japan Society for the Promotion of Science. This work was supported by the Grant-in-Aid for the Global COE Program ``The Next Generation of Physics, Spun from Universality and Emergence'' from the Ministry of Education, Culture, Sports, Science and Technology (MEXT) of Japan. This work was also supported by Grant-in-Aid of the MEXT No.\,18204015 and 20340043. 
\\
\\


\begin{thebibliography}{}
 
\bibitem[Anders \& Grevesse(1989)]{anders98} Anders, E., \& Grevesse, N.\ 1989, \gca, 53, 197 

\bibitem[Dahmen et al.(1998)]{dahmen98} Dahmen, G., Huttemeister, S., Wilson, T.~L., \& Mauersberger, R.\ 1998, \aap, 331, 959 

\bibitem[Ebisawa et al.(2008)]{ebi08} Ebisawa, K., et al.\ 
2008, \pasj, 60, 223 

\bibitem[Inui et al.(2009)]{inui09} Inui, T., Koyama, K., Matsumoto, H., \& Tsuru, T.~G.\ 2009, \pasj, 61, 241 

\bibitem[Kaastra \& Mewe(1993)]{kaa93} Kaastra, J.~S., \& Mewe, R.\ 1993, \aaps, 97, 443 

\bibitem[Koyama et al.(1989)]{koyama89} Koyama, K., Awaki, H., 
Kunieda, H., Takano, S., \& Tawara, Y.\ 1989, \nat, 339, 603 

\bibitem[Koyama et al.(1996)]{koyama96} Koyama, K., Maeda, Y., Sonobe, T., Takeshima, T., Tanaka, Y., \& Yamauchi, S.\ 1996, \pasj, 48, 249 

\bibitem[Koyama et al.(2007a)]{koyama07a} Koyama, K., et al.\ 2007a, \pasj, 59, S23 

\bibitem[Koyama et al.(2007b)]{koyama07b} Koyama, K., et al.\ 2007b, \pasj, 59, S221

\bibitem[Koyama et al.(2007c)]{koyama07c} Koyama, K., et al.\ 2007c, \pasj, 59, S245 

\bibitem[Koyama et al.(2008)]{koyama08} Koyama, K., Inui, T., Matsumoto, H., \& Tsuru, T.~G.\ 2008, \pasj, 60, S201 

\bibitem[Koyama et al.(2009)]{koyama09} Koyama, K., Takikawa, Y., Hyodo, Y., Inui, T., Nobukawa, M., Matsumoto, H., \& Tsuru, T.~G.\ 2009, \pasj, 61, 255 

\bibitem[Kushino et al.(2002)]{kushino02} Kushino, A., Ishisaki, 
Y., Morita, U., Yamasaki, N.~Y., Ishida, M., Ohashi, T., 
\& Ueda, Y.\ 2002, \pasj, 54, 327 

\bibitem[Masui et al.(2009)]{masui09} Masui, K., Mitsuda, K., Yamasaki, N.~Y., Takei, Y., Kimura, S., Yoshino, T., 
\& McCammon, D.\ 2009, \pasj, 61, 115 

\bibitem[Mitsuda et al.(2007)]{mitsuda07} Mitsuda, K., et al.\ 2007, \pasj, 59, S1 

\bibitem[Morris \& Serabyn(1996)]{moris96} Morris, M., \& Serabyn, E.\ 1996, \araa, 34, 645 

\bibitem[Mori et al.(2008)]{mori08} Mori, H., Tsuru, T.~G., Hyodo, Y., Koyama, K., \& Senda, A.\ 2008, \pasj, 60, 183 

\bibitem[Muno et al.(2006)]{muno06} Muno, M.~P., Bauer, F.~E., Bandyopadhyay, R.~M., \& Wang, Q.~D.\ 2006, \apjs, 165, 173 

\bibitem[Nakanishi \& Sofue(2003)]{nakanishi03} Nakanishi, H., \& Sofue, Y.\ 2003, \pasj, 55, 191 

\bibitem[Nakajima et al.(2009)]{nakajima09} Nakajima, H., Tsuru, T.~G., Nobukawa, M., Matsumoto, H., Koyama, K., Murakami, H., Senda, A., \& Yamauchi, S.\ 2009, \pasj, 61, 233

\bibitem[Nobukawa et al.(2008)]{nobu08} Nobukawa, M., et al.\ 2008, \pasj, 60, S191

\bibitem[Nobukawa et al.(2009)]{nobu09} Nobukawa, M., Koyama, 
K., Matsumoto, H., \& Tsuru, T.~G.\ 2009, \pasj, 61, 93

\bibitem[Rieke et al.(1989)]{reike89} Rieke, G.~H., Rieke, 
M.~J., \& Paul, A.~E.\ 1989, \apj, 336, 752 

\bibitem[Sakano et al.(2002)]{sakano02} Sakano, M., Koyama, K., Murakami, H., Maeda, Y., \& Yamauchi, S.\ 2002, \apjs, 138, 19 


\bibitem[Sawada et al.(2004)]{sawada04} Sawada, T., Hasegawa, 
T., Handa, T., \& Cohen, R.~J.\ 2004, \mnras, 349, 1167

\bibitem[Serlemitsos et al.(2007)]{ser07} Serlemitsos, P.~J., 
et al.\ 2007, \pasj, 59, S9 

\bibitem[Smith et al.(2001)]{smith01} Smith, R. K., et al.\ 2001, \apjs, 556, L91

\bibitem[Sofue(1995)]{sofue95} Sofue, Y.\ 1995, \pasj, 47, 527 

\bibitem[Tsuboi et al.(1999)]{tsuboi99} Tsuboi, M., Handa, T., 
\& Ukita, N.\ 1999, \apjs, 120, 1

\bibitem[Tsuru et al.(2009)]{tsuru09} Tsuru, T.~G., Nobukawa, M., Nakajima, H., Matsumoto, H., Koyama, K., \& Yamauchi, S.\ 2009, \pasj, 61, 219 

\bibitem[Yamauchi et al.(1990)]{yamauchi90} Yamauchi, S., Kawada, 
M., Koyama, K., Kunieda, H., \& Tawara, Y.\ 1990, \apj, 365, 532 

\bibitem[Yusef-Zadeh et al.(1999)]{yusef99} Yusef-Zadeh, F., 
Choate, D., \& Cotton, W.\ 1999, \apjl, 518, L33 

\end{thebibliography}
\end{document}